\documentclass[]{spie}  

\usepackage{amsmath,amsfonts,amssymb}
\usepackage{graphicx}
\usepackage[colorlinks=true, allcolors=blue]{hyperref}

\title{Spec-S5: A Next-Generation All-Sky Spectroscopic Facility Enabling Large-Scale Surveys for Cosmology and Astrophysics}

\author[a]{Claire Poppett}
\affil[a]{\small UC Berkeley Space Sciences Lab, Berkeley, CA 94520} 
\author[b]{David J. Schlegel}
\affil[b]{\small Lawrence Berkeley National Lab, 1 Cyclotron Rd, Berkeley, CA 94720, USA}
\author[c]{Arjun~Dey}
\affil[c]{\small NSF NOIRLab, 950 N. Cherry Avenue, Tucson, AZ 85719, USA}
\author[d]{Klaus~Honscheid}
\affil[d]{\small The Ohio State University, 191 West Woodruff Avenue, Columbus, OH 43210, USA}
\author[d]{Paul~Martini}
\author[e,f,g]{Alex~Drlica-Wagner}
\affil[e]{\small Fermi National Accelerator Laboratory, P.O.\ Box 500, Batavia, IL 60510, USA}
\affil[f]{\small Department of Astronomy \& Astrophysics, Chicago, IL 60637, USA}
\affil[g]{\small Kavli Institute of Cosmological Physics, Chicago, IL 60637, USA}
\author[a]{Pat Jelinsky}
\author[h]{Chang-Jin Oh}
\affil[h]{\small Wyant College of Optical Sciences, The University of Arizona, AZ 85721}
\author[i]{Konstantina Boutsia}
\affil[i]{\small Cerro Tololo Inter-American Observatory, NSF NOIRLab, Chile}
\author[b]{Joseph H. Silber}
\author[b]{Nicholas R. Wenner}

\authorinfo{Further author information: (Send correspondence to Claire Poppett)\\Claire Poppett: E-mail: clpoppett@berkeley.edu}

\begin{document} 
\maketitle

\begin{abstract}
The Stage-5 Spectroscopic Experiment (Spec-S5) is a next-generation, all-sky spectroscopic facility designed to address fundamental questions in cosmology and astrophysics. Building on the legacy of the Dark Energy Spectroscopic Instrument (DESI), Spec-S5 will upgrade two existing 4-m telescopes into 6-m, wide-field observatories, each equipped with a highly multiplexed spectrograph capable of measuring 13,000 spectra simultaneously. This overview paper summarizes the science motivation, system architecture, and integration strategy of the project. Spec-S5 will deliver a more than tenfold increase in spectroscopic capability, enabling transformative surveys in the post-Rubin, post-DESI era and advancing our understanding of dark matter, dark energy, and cosmic structure formation. 
\end{abstract}


\section{Introduction}

Wide-field spectroscopic surveys have transformed observational cosmology and astrophysics over the past several decades. Early efforts such as the CfA Redshift Survey, the Las Campanas Redshift Survey, and the 2dF Galaxy Redshift Survey established the power of large redshift catalogs for studies of cosmic structure \cite{Geller1989,Shectman1996,Colless2001}. Building on this foundation, the Sloan Digital Sky Survey (SDSS), followed by the Baryon Oscillation Spectroscopic Survey (BOSS)\cite{Dawson2013}, eBOSS\cite{Dawson2016}, and DESI\cite{Adame2025}, enabled increasingly precise measurements of large-scale structure, baryon acoustic oscillations (BAO), redshift-space distortions (RSD), galaxy evolution, and the structure of the Milky Way. The next decade will bring a dramatic increase in imaging survey capability through facilities such as the Vera C. Rubin Observatory\cite{LSST2019}, Euclid\cite{Euclid2011}, and the Nancy Grace Roman Space Telescope\cite{Roman2022}. These imaging surveys will identify billions of astronomical sources and transient events, creating a critical need for corresponding spectroscopic follow-up at unprecedented scale.

The Stage-5 Spectroscopic Experiment (Spec-S5) is a proposed next-generation wide-field spectroscopic facility designed to enable Stage-5 cosmology and a broad range of astrophysical investigations. The scientific case is closely aligned with priorities identified by the Astro2020 Decadal Survey and the Particle Physics Project Prioritization Panel (P5), positioning Spec-S5 as a potential flagship facility for the post-DESI era.

The baseline concept upgrades two existing 4-m class observatories---the Mayall Telescope at Kitt Peak National Observatory and the Blanco Telescope at Cerro Tololo Inter-American Observatory---into dedicated 6-m wide-field spectroscopic facilities operating in both hemispheres. Each observatory is designed to support a field of view exceeding two degrees in diameter and a focal-plane instrument capable of simultaneously observing approximately 13,000 targets. The baseline facility parameters are given in table \ref{tab:S5_baseline}.

\begin{table}[ht]
\centering
\caption{Baseline Spec-S5 facility parameters.}
\begin{tabular}{lc}
\hline
Parameter & Value \\
\hline
Primary mirror diameter & 6.0 m \\
Field of view & 2.2 deg \\
Fibers per observatory & 13,000 \\
Spectrographs & 23 \\
Wavelength coverage & 360--980 nm \\
Survey area (wide) & 25,000 deg$^2$ \\
Survey area (deep) & 11,000 deg$^2$ \\
Survey duration & 6 years \\
Redshift reach & $z\sim4.5$ \\
\hline
\end{tabular}
\label{tab:S5_baseline}
\end{table}

Spec-S5 is intended to provide an all-sky spectroscopic capability with substantially greater survey speed, multiplexing, and sensitivity than current facilities. Its primary science goals include precision measurements of large-scale structure, primordial inflation, dark energy, the neutrino mass, and dark matter, together with transformational datasets for studies of galaxy evolution, Galactic archaeology, and time-domain astronomy \cite{Besuner2025,Ferraro2022}.

A central feature of the project is the strategic reuse of existing observatory infrastructure and mature spectroscopic technologies. This approach minimizes technical risk, reduces development cost, and provides a practical path toward rapid deployment of a Stage-5 spectroscopic facility.

This paper provides an overview of the scientific motivation, observatory architecture, instrumentation concept, survey strategy, and systems integration approach for Spec-S5. Section \ref{sec:science} summarizes the primary science drivers motivating the facility. Section \ref{sec:platforms} describes the observatory and telescope architecture. Section \ref{sec:instr} reviews the instrument design. Section \ref{sec:survey} discusses survey operations. Section \ref{sec:system} outlines systems engineering and integration considerations.

\section{Science motivation}
\label{sec:science}

Spec-S5 is designed to address major open questions in cosmology and astrophysics through a transformational increase in spectroscopic survey capability. Building on the scientific legacy of SDSS, BOSS, eBOSS , and DESI, the facility will provide all-sky spectroscopy with unprecedented multiplexing and survey speed.

The primary cosmological objective is the precision mapping of large-scale structure over a broad redshift range that extends to $z \sim 4.5$. These measurements will enable improved constraints on dark energy --- from baryon acoustic oscillations and redshift-space distortions --- and on the neutrino mass and primordial inflation. In particular, the large survey volume and dense high-redshift sampling are designed to probe inflationary physics from primordial non-Gaussianity and other features in the primordial power spectrum.

\begin{figure}[thb]
    \centering
    \includegraphics[width=6.3in, trim={0 0.3cm 0 0}, clip]{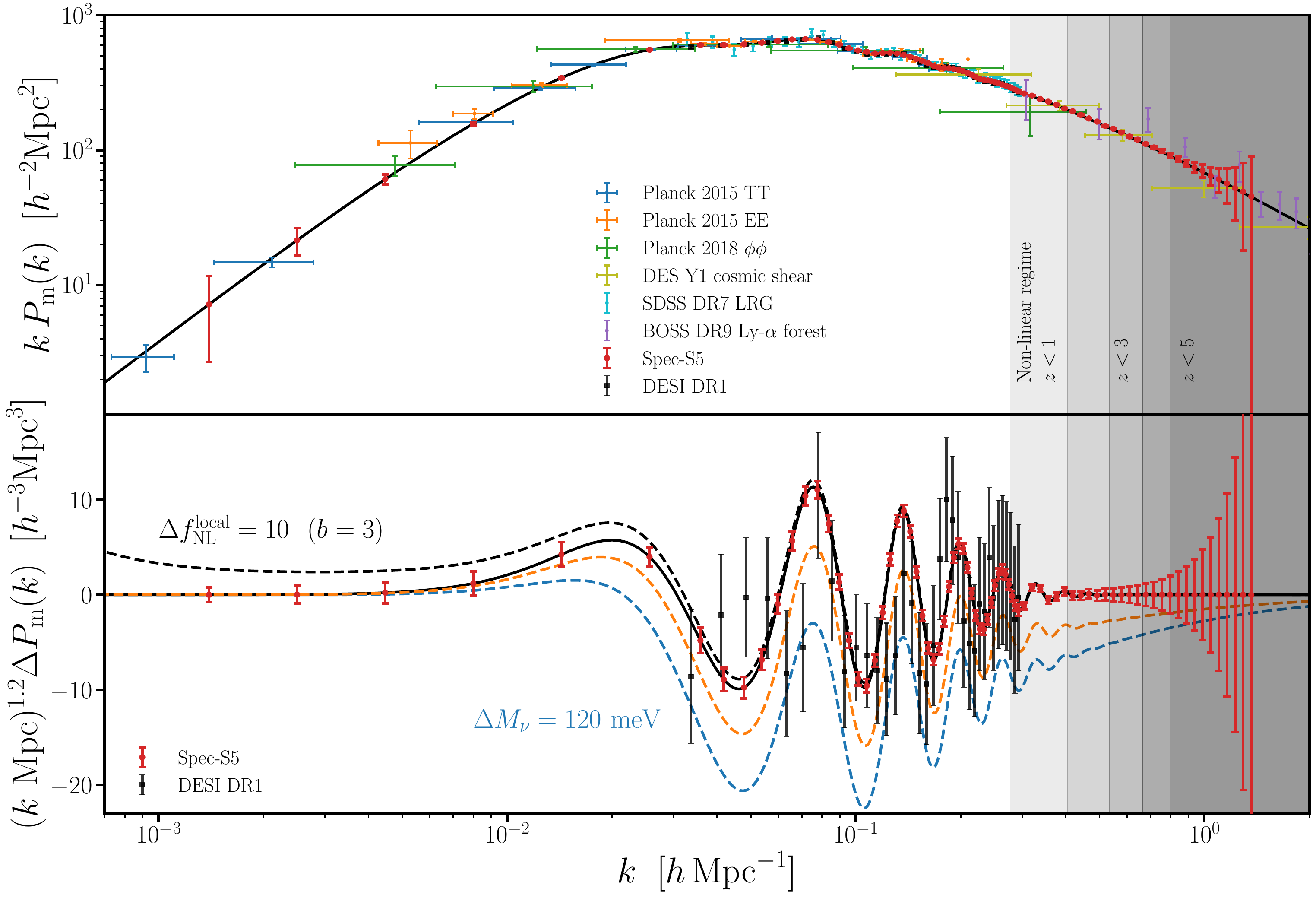}
    \caption{
    {\bf Top:}  Measurements of the power spectrum resulting from a wide array of CMB and galaxy redshift surveys. Spec-S5 (red points) represents a significant improvement on existing (and planned) measurements, resulting in very precise constraints across a wide range of $k$. 
    {\bf Bottom:}  The normalized power spectrum, showing the small deviations that result from the presence of non-Gaussianity (black dashed line), a massive neutrino (blue dashed line), or light relic particles (orange dashed line).  Spec-S5 has the potential to measure these small deviations out to $k\approx0.7$ in the matter-dominated, high-redshift universe.
    }
    \label{fig:powerspectrum}
\end{figure}

Figure~\ref{fig:powerspectrum} shows power-spectrum measurements from Cosmic Microwave Background (CMB) and redshift survey experiments. Deviations from Gaussianity, changes in the number of relativistic species, or differences in the sum of the neutrino masses are hard to distinguish with current experiments, as shown by the curves in the bottom panel. Spec-S5 will have the sensitivity to discern these effects over a wide range of scales, as demonstrated by the red points, which show forecast uncertainties for a Spec-S5 survey of 62 million galaxies in the redshift range $2.1 < z < 4.5$. 

Another way to quantify the performance of Spec-S5 is through the concept of a primordial figure of merit (PFoM), introduced by Sailer et al.\cite{sailer21}, which maps approximately linearly to improved %
constraints on these fundamental effects. The baseline high-redshift survey with Spec-S5 has a PFoM of 9--10. This represents a 10$\times$ improvement over the current state-of-the-art (DESI; PFoM $\sim 0.9$) and the ESA Euclid space mission (PFoM $\sim 1$), both of which use smaller telescopes to survey a smaller volume of the low-redshift universe, with correspondingly less sensitivity to primordial physics.

Spec-S5 will also support major advances in understanding dark matter and galaxy evolution. Radial velocity measurements for tens of millions of stars, combined with Gaia and Rubin astrometry, will probe Galactic halo structure and dark matter substructure. Simultaneously, spectra for more than one hundred million galaxies and quasars will enable statistical studies of galaxy assembly, chemical enrichment, feedback, and environmental dependence across most of cosmic history.

The facility is strongly synergistic with Rubin Observatory, Euclid, and Roman. Rubin imaging provides the photometric foundation for target selection and transient discovery, while Spec-S5 spectroscopy supplies precise redshifts and astrophysical characterization at the scale required for Stage-5 cosmology.

\section{Observatory architecture: Dual-Hemisphere Strategy}
\label{sec:platforms}

The baseline Spec-S5 architecture consists of twin wide-field spectroscopic observatories operating in the northern and southern hemispheres, providing access to nearly the entire extragalactic sky. This dual-hemisphere strategy maximizes the accessible cosmological volume and enables extensive overlap with major imaging and multi-wavelength surveys of the 2030s. By upgrading existing observatory infrastructure rather than constructing new facilities, Spec-S5 offers a cost-effective and comparatively low-risk path toward an all-sky Stage-5 spectroscopic program.

\begin{figure}[ht]
    \centering
    \begin{tabular}{cc}
        \includegraphics[width=0.33\textwidth]{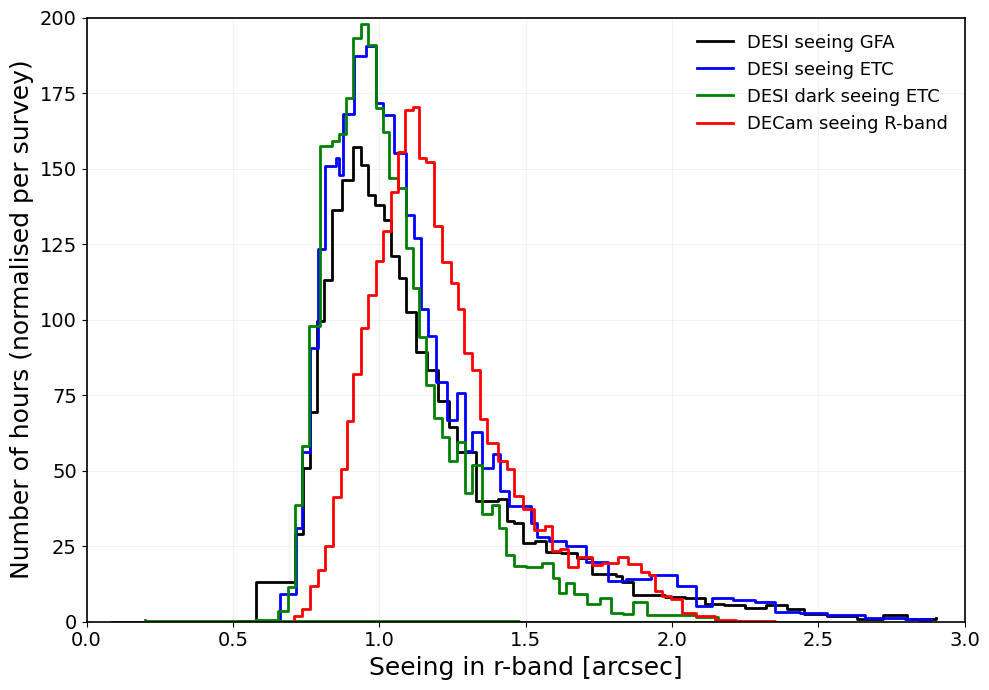} &
        \includegraphics[width=0.6\textwidth]{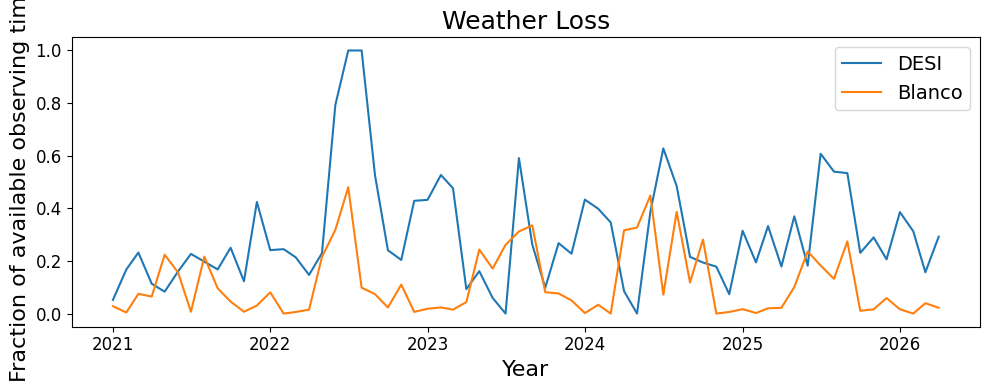} \\
    \end{tabular}
    \caption{Since we have operated both DESI and DECam for multiple years, we have excellent data on site conditions. The panel on the left shows the distributions of seeing for DESI and DECam using all available data. The median seeing in the DESI data is 1.05 arcsec. The panel on the right shows the time lost due to weather during the DESI survey and at Blanco site during the same time period.}
    \label{fig:seeing}
\end{figure}

We have excellent long-term data for evaluating both the Mayall and Blanco sites as shown in figure \ref{fig:seeing}.  The night-time sky brightness, atmospheric transmission and atmospheric transparency (cloudiness) distributions between the two sites are nearly identical.  We characterize these sites by the number of available ``effective hours,'' which is the time equivalent to observing during nominal, photometric, dark-time conditions. Nominal conditions are defined as a sky brightness of 21.07 mag/arcsec$^2$, seeing of 1.10 arcsec FWHM in $r$-band, and 100\% transparency. At each telescope, there are 650 effective hours per year when dark-time targets may be observed and 65 effective hours per year when bright-time targets may be observed. The baseline survey for Spec-S5 is predicted to take 6 years, resulting in 100 million effective fiber hours of dark-time observing and 10 million effective fiber hours of bright-time observing.

\section{Instrument design}
\label{sec:instr}
The Spec-S5 instrument concept combines a high-density robotic focal plane, a wide-field optical system, and replicated fiber-fed spectrographs to deliver unprecedented spectroscopic survey throughput. The instrument architecture is organized into modular subsystems that simplify fabrication, integration, testing, and long-term maintenance while providing a scalable path toward the multiplexing required for Stage-5 cosmology. Each observatory will support approximately 13,000 simultaneously deployed fibers feeding an array of replicated spectrographs operating over the wavelength range 360--980\,nm. Together, these systems provide the sensitivity, survey speed, and operational efficiency needed to execute the Spec-S5 science program. The following subsections detail the focal plane, optical design, and spectrograph architecture of the Spec-S5 instrument.

\subsection{Focal Plane Design}
The focal plane integrates robotic fiber positioners, guide and focus cameras, fiducial illumination sources, and calibration fibers into a modular architecture optimized for maintainability and operational reliability. The baseline design contains approximately 13,000 robotic fiber positioners distributed across the focal surface at a pitch of 6.2\,mm, substantially denser than the DESI focal plane, which packs 5000 fibers with a pitch of 10.4\,mm.

The positioners are organized into modular ``rafts,''\cite{SilberRobotsASPE2022} each containing 63 fiber actuators and associated electronics. This approach simplifies integration, servicing, and replacement while reducing operational downtime. The modular architecture also supports future upgrades in fiber-positioning technology.

The robotic actuators are designed for rapid reconfiguration and sub-arcsecond positioning accuracy. The focal plane system includes closed-loop metrology and calibration hardware to ensure accurate fiber placement across the full field of view. The resulting multiplex advantage is a central enabling capability for the Spec-S5 survey strategy.

The final design of the robotic fiber positioner system has not yet been baselined, although multiple promising technologies are currently under active development and evaluation. Existing R\&D programs have already demonstrated compact actuator architectures capable of meeting the required packing density, positioning accuracy, reliability, and reconfiguration speed needed for Spec-S5. Several candidate designs build directly on the technological heritage of DESI while incorporating advances in miniaturized motors, control electronics, metrology, and modular focal-plane integration. The maturity of these parallel development efforts substantially reduces technical risk and provides flexibility in optimizing the final positioner architecture for performance, manufacturability, and long-term operational robustness. \cite{Wenner2026,Schubnell2026}.

\subsection{Optical Design}
\begin{figure}[thb]
    \centering
    \includegraphics[width=0.8\textwidth]{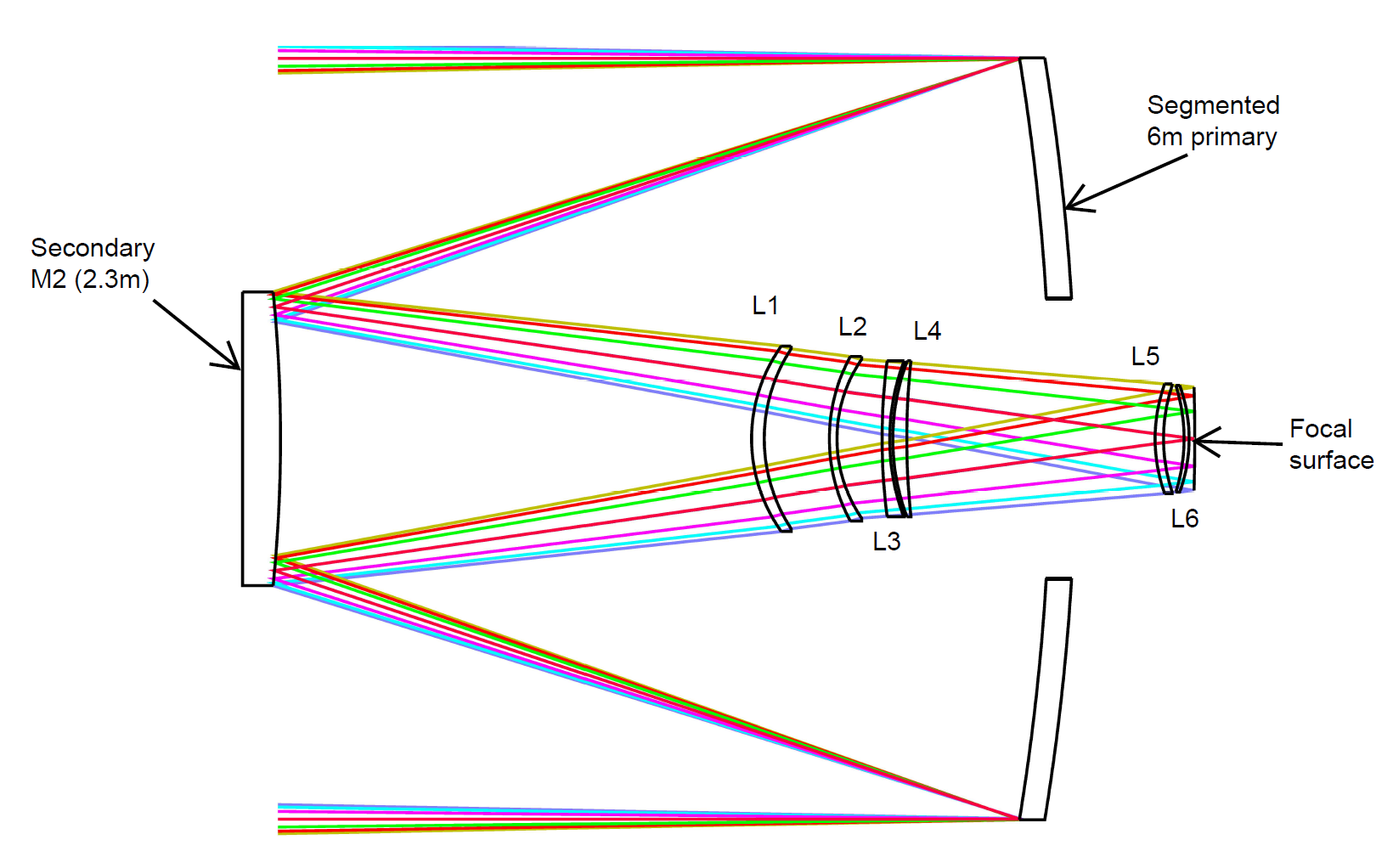}
    \caption{Optical design of the telescope mirrors and corrector. All corrector lenses are fused silica with spherical surfaces and the design meets baseline requirements for Spec-S5. L3 and L4 are the ADC; they rotate to remove atmospheric dispersion up to 60$^\circ$ from zenith. The focal plane diameter is 818.4\,mm with a 2.17$^\circ$ FOV.}
    \label{fig:optics}
\end{figure}

The baseline optical configuration for Spec-S5, shown in figure \ref{fig:optics}, is a wide-field Cassegrain design consisting of a 6.0-m primary mirror, a 2.3-m secondary mirror, and a six-element corrector delivering a corrected field of view that is approximately $2.2^{\circ}$ in diameter. The optical system is designed to support a focal plane diameter of 0.82 m at $f/3.6$, providing excellent matching to the fiber-fed spectrographs while maintaining high throughput across the full field.

The optical prescription was developed through a trade study emphasizing survey speed, manufacturability, and operational simplicity. A Cassegrain configuration offers several advantages for Spec-S5, including balanced telescope loading and shorter fiber runs between the focal plane and spectrographs compared to prime-focus architectures. Reduced fiber length improves blue-wavelength throughput, which is particularly important for optimizing observations of high-redshift galaxies and the Ly$\alpha$ forest.

The corrector assembly consists entirely of fused-silica lenses with spherical surfaces, simplifying fabrication relative to systems requiring large aspheric optics. Two counter-rotating lenses act as an atmospheric dispersion corrector (ADC), enabling efficient observations to zenith angles of $60^{\circ}$. The design maintains excellent image quality and low chief-ray deviation across the full field, with focal-ratio variation of $\sim$2\%, which is well below the baseline requirement for Spec-S5 of 1\%.

\begin{figure}[ht]
    \centering
    \begin{tabular}{cc}
        \includegraphics[width=0.48\textwidth]{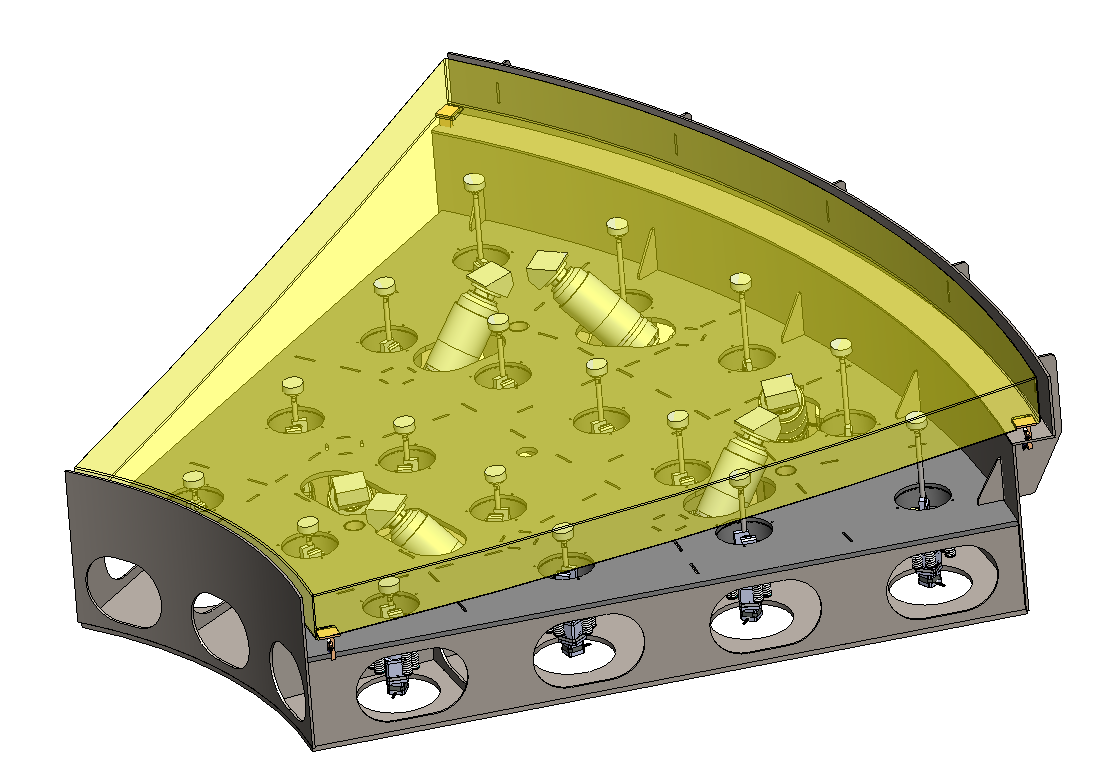} &
        \includegraphics[width=0.48\textwidth]{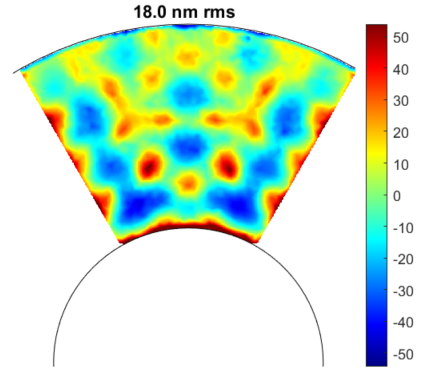} \\
    \end{tabular}
    \caption{Each segment of the Spec-S5 120-mm thick meniscus mirror can be supported and controlled with a simple arrangement of 17 active actuators per segment.}
    \label{fig:mirror}
\end{figure}

A recent optical design study, funded by the Heising-Simons Foundation, was conducted by the College of Optical Sciences at the University of Arizona. This trade study evaluated multiple primary mirror configurations, including substrate type, segment count, and support strategies. Quantitative comparisons of cost, manufacturability, actuator count, blank size, and structural performance led to the selection of an 8-segment, 120-mm thick meniscus design, each segment with 17-point active axial support and bipod lateral supports. This design is shown on the left-hand side of figure \ref{fig:mirror}. High-fidelity finite element analysis (FEA) was performed as shown in the right panel of figure \ref{fig:mirror} and predicts 15.6\,nm RMS by support print through and polishing combined, $\sim$10\,nm at the horizon under gravity loading, and $\sim$2.1\,nm RMS/$^\circ$C thermal distortion—within system allocations—indicating readiness for prototype fabrication.

Preliminary work on the secondary mirror, telescope structure, metrology, and wavefront sensing has also progressed. The preferred design for the M2 Secondary mirror (M2) is a 100-mm thick meniscus with 27 axial supports plus 6 lateral supports. This design meets the $\leq$30\,nm RMS surface-figure requirement, and metrology approaches based on convex mirror subaperture stitching interferometry have been defined. Structural analyses confirm acceptable gravity-induced alignment behavior, while a hybrid wavefront-sensing approach combining phase retrieval and dispersed fringe sensing has been validated through simulation and tests.

Finally, it should also be noted that FEA of the horseshoes for both the Mayall and the Blanco telescopes demonstrates that the mount can be modified to accommodate a 6-meter-class primary mirror without compromising structural integrity. 

Together, the results outlined in this subsection establish a coherent optical–mechanical concept, analytically validate performance, and define alignment, commissioning, and risk-mitigation pathways for Spec-S5 upgrades. 

\subsection{Spectrograph design}
Spec-S5 uses replicated fiber-fed spectrographs derived from the highly successful DESI design. Each spectrograph splits incoming light into three wavelength channels covering approximately 360--980\,nm, with spectral resolution ranging from $R \sim 2000$ in the blue to $R \sim 5500$ in the near-infrared.

Each observatory will host 23 spectrographs fed by approximately 13,000 fibers. The spectrographs are housed within a thermally stabilized enclosure to ensure calibration stability and high-quality sky subtraction for faint-source spectroscopy.

A major enhancement relative to DESI is the planned deployment of low-noise multi-amplifier sensing (MAS) CCDs. These detectors leverage developments from Skipper CCD technology to achieve sub-electron read noise while maintaining practical readout speeds. The reduced read noise is particularly important for spectroscopy of faint blue galaxies at $z\leq2$ where detector noise can otherwise dominate the signal budget.

The instrument design emphasizes high throughput, operational stability, and scalability while minimizing technical risk through extensive reuse of proven DESI technologies.

\section{Survey design and operations}
\label{sec:survey}

\begin{figure}[htbp]
    \centering
    \includegraphics[width=0.9\textwidth]{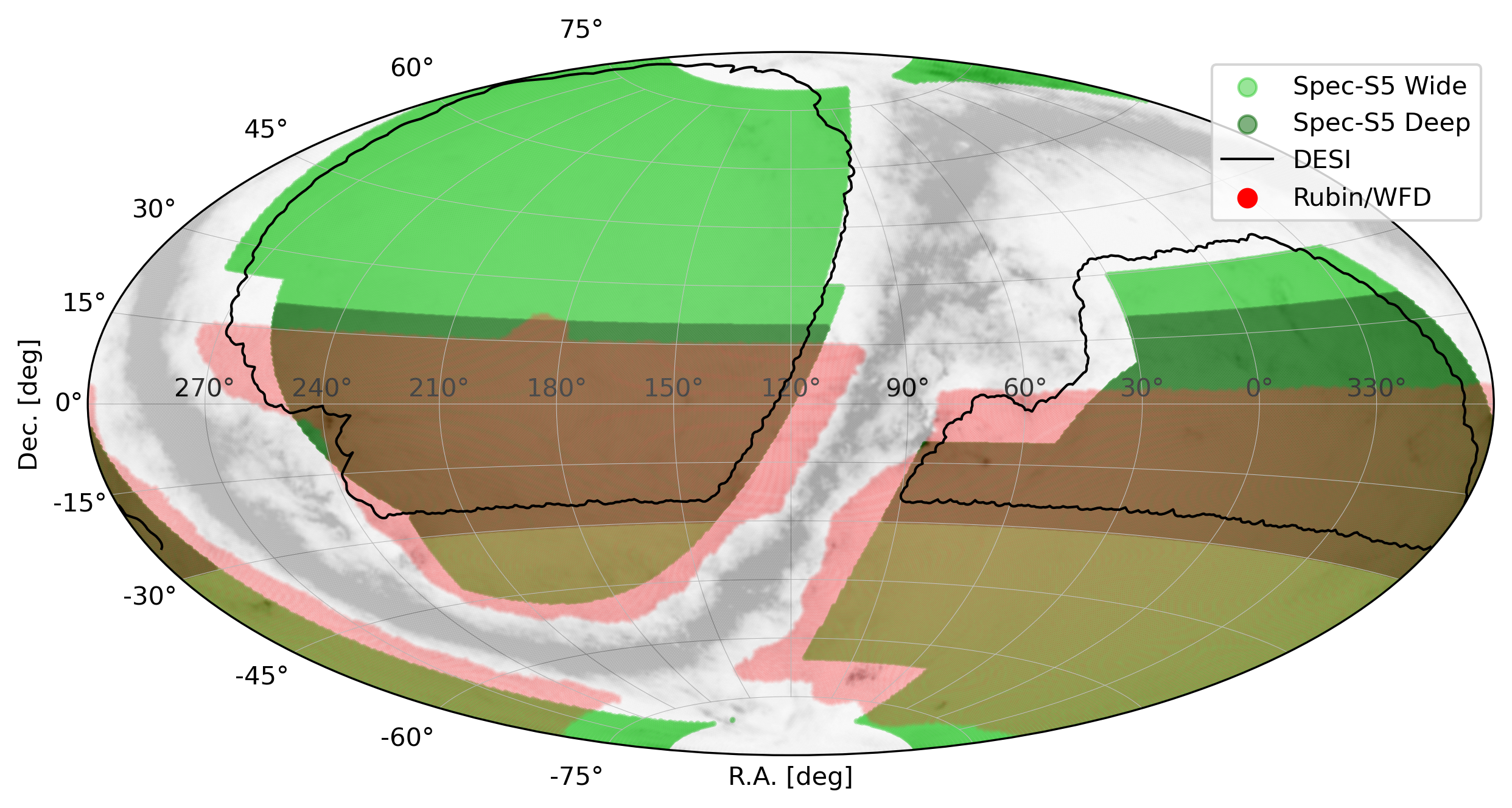}
    \caption{\label{fig:footprint}
    The footprints of the Spec-S5 Wide-Area Cosmology Survey (light green) and the Deep Survey of high-redshift galaxies (dark green) are shown, compared with the nominal footprints of the ongoing DESI spectroscopic survey (black line) and the Rubin Legacy Survey of Space and Time (pink).}
\end{figure}

The baseline survey strategy for Spec-S5 consists of a 25,000\,deg$^2$ wide-area cosmology survey together with an approximately 11,000\,deg$^2$ deep survey optimized for high-redshift structure formation studies. The survey footprint shown in figure \ref{fig:footprint} is selected to maximize overlap with Rubin LSST and other major cosmological facilities while avoiding regions of high Galactic extinction.

Spec-S5 observations in dark time will target luminous red galaxies, emission-line galaxies, quasars, Lyman-alpha forest quasars, Lyman-break galaxies, and Lyman-alpha emitters. Bright-time observations will focus on bright galaxies and stellar populations within the Milky Way. Combined, these surveys will produce spectra for hundreds of millions of objects over the operational lifetime of Spec-S5.

Operations will follow a dynamic observing strategy similar to DESI \cite{DESI2019}, in which exposure times are adjusted in real time according to observing conditions and target requirements. Rapid focal plane reconfiguration and automated scheduling will maximize observing efficiency.

The data management system is designed to process approximately 500,000 spectra nightly. Automated pipelines will perform image pre-processing, spectral extraction, calibration, redshift determination, and quality assessment. Open-source software practices and scalable computing infrastructure will be a key feature of Spec-S5 to ensure long-term survey operations and first-rate community access.

\section{Systems engineering and integration}
\label{sec:system}

Spec-S5 emphasizes the modularity, replication, and reuse of proven technologies to minimize technical risk and accelerate deployment. The project architecture divides the observatory into major subsystems including telescope optics, focal plane instrumentation, spectrographs, calibration systems, control software, and data systems.

A major advantage of the project is the extensive reuse of infrastructure and operational experience from DESI and existing observatory platforms. Nevertheless, targeted development is required for high-density fiber positioning, active support of the segmented primary mirrors, large-scale integration, and high-throughput data systems.

Integration and commissioning activities will be carefully staged to validate critical subsystems prior to full observatory operations. Early survey validation observations will be used to refine target selection, calibration procedures, and observing strategy before the start of full survey operations.

\subsection{Pathfinder instruments}

An important intermediate step toward the full Spec-S5 facility could be the deployment of a wide-field multiplexed spectroscopic instrument on the Blanco telescope following the conclusion of DECam operations. One attractive concept is a hybrid instrument combining a wide-field, massively multi-object spectroscopic (MOS) capability with a deployable integral field unit (IFU) mode. The central element in the focal plane would be a large IFU surrounded by MOS mode fibers. In MOS mode, the instrument would provide an efficient southern-hemisphere spectroscopic follow-up facility for Rubin LSST targets, including transient sources, photometric-redshift calibration samples, galaxy evolution surveys, and cosmological large-scale structure measurements. Such a facility would address the growing need for highly efficient spectroscopic follow-up in the Rubin era while serving as a technological and operational pathfinder for Spec-S5 focal plane systems, survey operations, and data infrastructure. The IFU capability would enable spatially resolved spectroscopy for nearby galaxies and ultra-faint dwarf systems, providing a powerful probe of dark matter distributions, stellar populations, and dynamical structure on small scales. Together, these capabilities would establish a flexible and scientifically compelling precursor facility while reducing technical risk for the full Spec-S5 program.

%
%
%

\section{Conclusions}
\label{sec:conc}
Spec-S5 is a proposed all-sky spectroscopic facility designed to extend precision large-scale structure measurements into the high-redshift universe while enabling a broad range of astrophysical investigations. The baseline design combines twin 6-m observatories, wide-field optics, high-density robotic fiber positioning, and replicated spectrographs to provide more than an order-of-magnitude increase in spectroscopic capability relative to current facilities.

Ongoing studies indicate that the required telescope, optical, focal-plane, and detector technologies can be realized through a combination of mature designs and targeted development programs. The extensive reuse of existing observatory infrastructure and operational experience provides a practical path toward implementation while reducing technical and programmatic risk.

With its combination of sky coverage, multiplexing, and survey speed, Spec-S5 would provide a foundational spectroscopic resource for the astronomical community in the 2030s, complementing Rubin, Euclid, Roman, and other major observational facilities.

\acknowledgments

This work was supported in part by the Director, Office of Science, Office of High Energy Physics of the US Department of Energy under contract No.\,DE-AC02-05CH11231, and by the Laboratory Directed Research and Development (LDRD) program of Lawrence Berkeley National Laboratory.
LBNL and the University of Arizona were supported in part by the Heising-Simons Foundation.
Development of the Spec-S5 science case has been supported in part by the Kavli Foundation (PS-2025-GR-249-3072).

The authors acknowledge the contributions of the broader Spec-S5 collaboration and the many institutions supporting the development of next-generation spectroscopic survey facilities. This work builds upon the heritage of SDSS, BOSS, eBOSS, DES, and DESI.

\end{document}